\pdfoutput=1

\documentclass[11pt,]{article}

\usepackage{algorithm}
\usepackage{algorithmic}

\usepackage[table]{xcolor}
\usepackage[]{acl}
\usepackage{graphicx}
\usepackage{tabularx}
\usepackage{multirow}
\usepackage{amsfonts}
\usepackage{amssymb}
\usepackage{placeins}
\usepackage[normalem]{ulem}
\usepackage{booktabs} 
\usepackage{float}   
\usepackage{fontawesome5}
\usepackage{times}
\usepackage{latexsym}
\usepackage{amsmath} 
\usepackage{amsthm}

\usepackage{subcaption}

\newcommand{\update}[1]{\textcolor{black}{#1}}

\usepackage[T1]{fontenc}

\usepackage[utf8]{inputenc}
\usepackage{enumitem}
\setlist[itemize]{noitemsep, nolistsep}

\usepackage{microtype}

\usepackage{inconsolata}

\title{Computed, not Memorized: Bit-Factorized Identifiers for Efficient Generative Retrieval} 

\author{I-Fan Lin \\
  \fontsize{11pt}{12pt}\selectfont Leiden University \\
  \fontsize{11pt}{12pt}\selectfont Leiden,  The Netherlands \\
  \texttt{\fontsize{9pt}{10pt}\selectfont i.lin@liacs.leidenuniv.nl} \\\And
    Zhen Zhang \\
  \fontsize{11pt}{12pt}\selectfont Shandong University \\
  \fontsize{11pt}{12pt}\selectfont Qingdao, China \\
  \texttt{\fontsize{9pt}{10pt}\selectfont zhen.zhang.sdu@gmail.com} \\\And
  Faegheh Hasibi \\
  \fontsize{11pt}{12pt}\selectfont Radboud University \\
  \fontsize{11pt}{12pt}\selectfont Nijmegen, The Netherlands\\
  \texttt{\fontsize{9pt}{10pt}\selectfont faegheh.hasibi@ru.nl} \\\\\And
  Suzan Verberne \\
  \fontsize{11pt}{12pt}\selectfont Leiden University \\
  \fontsize{11pt}{12pt}\selectfont Leiden,  The Netherlands\\
\texttt{\fontsize{9pt}{10pt}\selectfont s.verberne@liacs.leidenuniv.nl} \\}

\begin{document}
\maketitle
\begin{abstract}
Generative retrieval is known for its efficient end-to-end indexing and retrieval. In particular, numeric Document Identifiers (DocIDs) are generally more efficient because they are shorter and generally fixed-length 
sequences over a small vocabulary, requiring fewer decoding steps during beam search. However, they tend to generalize poorly to unseen documents and must be learned during indexing-stage training. 
In this paper, we propose a numeric DocID design for generative retrieval that enables better adaptation to unseen documents without retraining.
\update{Our DocID design inherits the language model's knowledge: each DocID is a sequence of tokens factorized into bits, and each token is scored as the sum of its bits. Unlike existing numeric DocIDs, whose tokens carry meaning only through embeddings memorized during indexing training, each bit here derives its meaning from the model's representation. Therefore, the DocIDs of unseen documents remain scorable without retraining.}
Experiments on three benchmarks show that our DocID design substantially improves unseen-document retrieval while maintaining comparable performance on seen documents. It achieves over 20 percentage points higher Hit@10 on unseen documents and up to 7$\times$ lower inference latency than existing numeric DocID designs on NQ320K and TriviaQA, while performing only 3 percentage points  below weight-update methods on MS MARCO without additional training.
These results show that an appropriate DocID design can heavily reduce retraining costs when adapting generative retrievers to new  documents.
\end{abstract}

\section{Introduction}
\begin{figure}[t]
        \centering
        \includegraphics[width=\columnwidth]{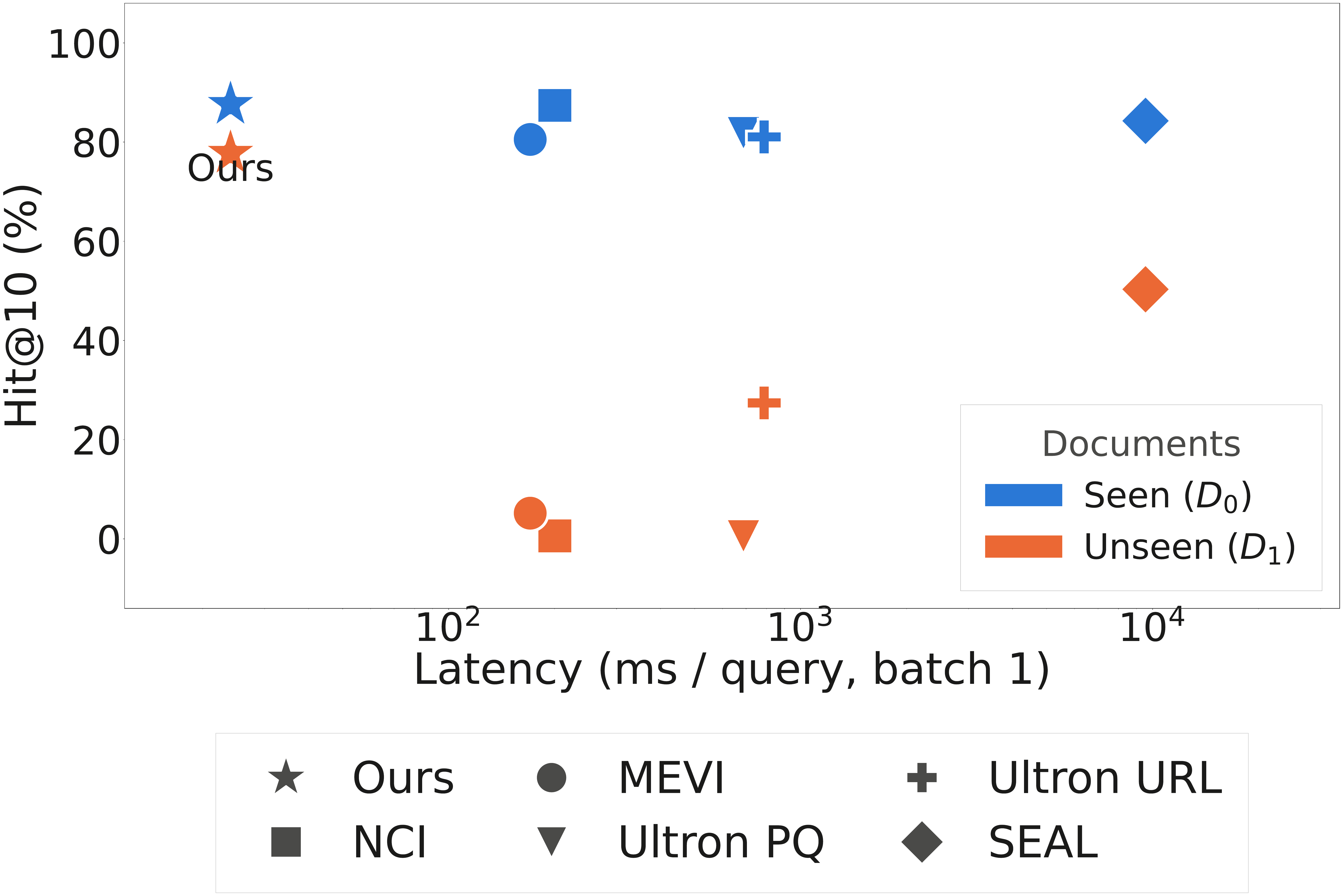}
        \caption{Our DocID design achieves comparable retrieval to the strongest baseline on seen documents, a clear margin on unseen documents, and the lowest latency of any method.}

        \label{fig:hit_10}
\end{figure}


Generative Retrieval is a retrieval paradigm in which an input query text is directly matched to a document identifier (DocID) using a language model, typically a sequence-to-sequence model. Unlike conventional dense retrieval methods, which require a vector database or Approximate Nearest Neighbour (ANN) index to store document embeddings and perform nearest-neighbor search at inference time, generative retrieval retrieves documents by generating their DocID directly \cite{li2025matching}. This design can offer inference and memory efficiency advantages \citep{kim2024exploring}.

Two different DocID designs exist for generative retrieval systems: numeric DocIDs and textual DocIDs. Both have their own strengths: Numeric DocIDs offer computational advantages, as they reduce the complexity of beam-search decoding \cite{ wang2022neural, sun2023learning,zhao2025unifying}. In contrast, textual DocIDs can exploit the semantic priors encoded in pretrained language models and are able to give more interpretable document identifiers \cite{bevilacqua2022autoregressive, zhou2022ultron, askari2026generative}.

One of the major challenges in current generative retrieval research is adapting to the retrieval of previously unseen documents. 
Generative retrieval models typically require an indexing stage during training, where the model learns to associate documents with their corresponding DocIDs. 
As a result, the model tends to be biased toward DocIDs seen during training. This lack of generalization is especially pronounced for numeric DocIDs \cite{zhang2025replication}, but also textual DocIDs lead to substantially lower effectiveness on unseen than on seen documents \citep{zhang2025replication}. 

While incremental training methods adapt numeric DocIDs to new documents \citep{mehta2023dsi++, kishore2023incdsi}, updating model parameters remains computationally expensive. Recent model-editing approaches update only a small subset of parameters while achieving performance comparable to full retraining \citep{zhang2026model}. However, editing the model for newly added documents still requires computation with each update, making frequent updates costly.

To address the poor adaptation of numeric DocIDs and the cost of updating model weights, we propose \textbf{Co}mputed \textbf{B}it-Factorized \textbf{I}dentifiers for \textbf{G}enerative \textbf{R}etrieval (\textbf{CoBIGR}). CoBIGR represents each document with a sign-projection binary code of its document vector and factorizes the retrieval score over its identifier bits, which inherit the model's knowledge of the document rather than a memorized embedding learned from indexing training. Additionally, the proposed bit-factorized IDs are flat and their score is permutation-invariant. This property enables multi-tree beam search during inference, which substantially improves effectiveness on unseen documents, while scoring with one forward pass keeps decoding efficient (see Figure \ref{fig:hit_10}). 

Our contributions are: (1) We propose a bit-factorized numeric DocID: its bits derive meaning from the language model, so unseen DocIDs can still be retrievable. (2) We study direct DocID adaptation to dynamic corpora without weight updates, addressing an underexplored gap in costly generative retrieval training. (3) Our experiments show that our approach achieves comparable performance on seen documents to prior DocID designs while substantially improving performance on unseen documents, with 40--60 percentage points higher Hit@1 and Hit@100 than other numeric DocIDs. It also remains within 3 percentage points of methods that update model weights. Our source code is available at \url{https://anonymous.4open.science/r/bit_factorized_token-2C13/}

\section{Related Work}

\paragraph{DocID in Generative Retrieval}
Generative retrieval maps a query directly to a DocID. During training, the DocID serves as the output target for both document-to-DocID (indexing) and query-to-DocID (retrieval) training. For text-based identifiers, titles \citep{de2020autoregressive}, URLs \citep{zhou2022ultron}, document substrings \citep{bevilacqua2022autoregressive}, and term sets \citep{zhang2024generative} are commonly used as identifiers. In contrast, numeric identifiers are often derived through using clustering methods such as k-means \citep{wang2022neural} or through representation-based quantization methods such as Product Quantization (PQ), Residual Quantization (RQ) and RQ-VAE \citep{zhou2022ultron, zhang2023mevi,rajput2023recommender}. These IDs are semantically structured. Although the numeric identifiers produce compact beam trees and enable more efficient decoding, they often suffer from limited generalization.   

\paragraph{Hashing in Retrieval}
Hashing has been widely used for efficient retrieval by providing compact representations and approximate similarity measures based on Hamming distance \citep{yamada2021efficient, he2025survey}. Recently, FLASH \citep{zhang2026flash} uses hash codes as the first step of retrieval: the generated code seeds a Hamming-radius search and is reconstructed into a continuous embedding for reranking. \update{Our method follows generative retrieval paradigm: it directly generates valid DocIDs, decoded token by token under a prefix-tree constraint and ranked by log-probability.}

\paragraph{Dynamic Corpus Adaptation for Numeric Identifiers}
Existing work primarily focuses on retraining the retrieval model when new documents are arrived: DSI++ \citep{mehta2023dsi++} reduces catastrophic forgetting by optimizing for flatter loss landscapes and incorporating pseudo-queries in continual training. IncDSI \citep{kishore2023incdsi} perform a constraint optimization to adapt to dynamic corpus, this avoids the full re-training of the decoder. However, the method is restricted to atomic document identifiers. DOME \citep{zhang2026model} further generalizes document insertion by optimizing edit vectors with soft-label guidance, enabling effective adaptation under structured document identifier schemes. Unlike prior retraining-based approaches, we study the zero-shot adaptation of newly arrived documents into structured numeric identifier spaces, which, to the best of our knowledge, has remained underexplored.

\section{Method}

\subsection{Problem Formulation}

\paragraph{Generative retrieval}

Let $D=\{d_1,\dots,d_N\}$ be a document corpus. Each document $d$ is associated with a DocID $\mathbf{c}_d=(c_1,\dots,c_T)$, a length-$T$ sequence of tokens from a predefined vocabulary. Given a query $q$, a generative retrieval model $p_\theta$ returns a ranked list of DocIDs by directly generating them.
Typically, $p_\theta$ is a sequence-to-sequence model that factorizes the DocID probability autoregressively,
\begin{equation}
  p_\theta(\mathbf{c}_d \mid q)=\prod_{t=1}^{L} p_\theta\!\left(c_t \mid c_{<t}, q\right),
\end{equation} The most relevant DocIDs are obtained by constrained beam search over the prefix tree of valid identifiers, so that only DocIDs of documents in $D$ can be produced. Training combines two objectives: an \emph{indexing} step that learns to bind each document $d$ to its DocID $\mathbf{c}_d$, and a \emph{retrieval} step that learns to map a query to the DocIDs of its relevant documents.

\paragraph{Dynamic Corpus}
In practice, the document collection $D$ changes over time. We therefore consider newly arrived documents after the model is trained. The model is trained on an initial seen document set $D_0$, using queries whose relevant documents all belong to $D_0$. After training, a set of new documents $D_1$ arrives, with $D_0 \cap D_1=\emptyset$, resulting in the expanded corpus $D_0\cup D_1$. We evaluate retrieval on the expanded corpus using two types of test queries: seen-document queries, whose relevant documents belong to $D_0$, and unseen-document queries, whose relevant documents belong to $D_1$.

\subsection{Proposed Approach}
\begin{figure*}[t]
        \centering
        \includegraphics[width=1\textwidth]{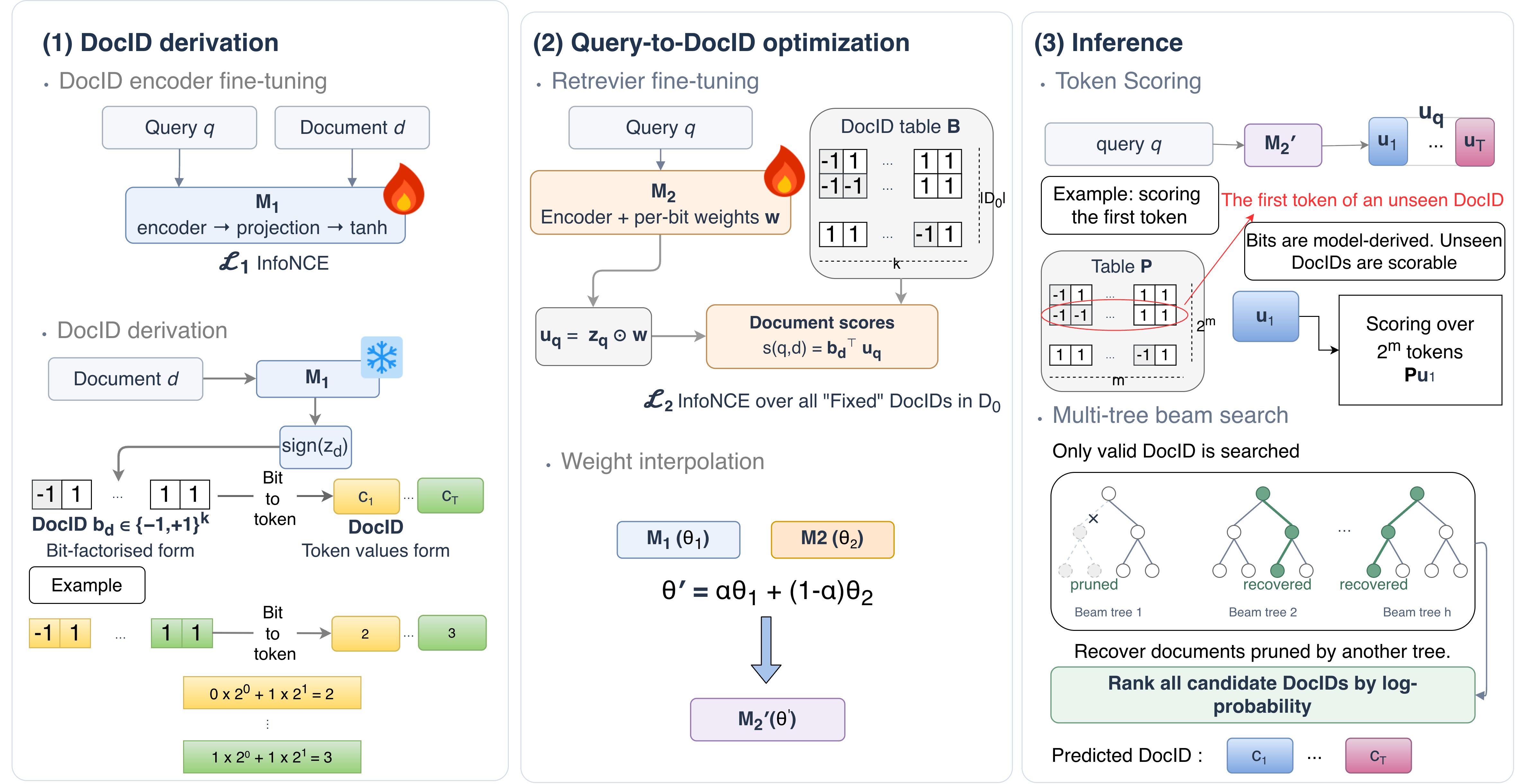}
        \caption{Overview of CoBIGR: (1) DocID derivation: Trains documents to map to separable binary codes, so each document derives its DocID through direct computation instead of assignment. (2) Query-to-DocID optimization: Trains the retriever to score gold DocIDs highly. Since training is based on seen documents, we reduce the resulting bias through weight interpolation. (3) Inference: We use lookup table to do the scoring. Since our DocID tokens are flat, we decode with constrained beam search over multiple permuted prefix trees.}
        \label{fig:pipeline}
\end{figure*}

Our proposed pipeline is summarized in Figure \ref{fig:pipeline}.

\subsubsection{DocID derivation}

To ensure DocIDs inherit the knowledge of a pretrained language model (PLM) encoder, we derive them directly from the PLM output. Given the encoder embedding $\mathbf{e}$ of a text, we compute
\begin{equation}\label{archi}
  \mathbf{z} = \tanh\big(\mathrm{LN}(\mathbf{W}\mathbf{e})\big) \in (-1,1)^k
\end{equation}
where $\mathbf{W} \in \mathbb{R}^{k \cdot l}$ is a projection layer that decouples the DocID length $k$ from the encoder's hidden size $l$. $\mathrm{LN}$ is layer normalization used to stabilize the scale~\citep{ba2016layer}. Following \citet{yamada2021efficient}, the tanh bounds each dimension so that its sign can serve as a bit: a document's DocID bit vector is $\mathbf{b}_d = \operatorname{sign}(\mathbf{z}_d) \in \{-1,+1\}^k$. The entire model is trained with InfoNCE contrastive loss \citep{oord2018representation, karpukhin2020dense} between query and document representations at the tanh output:
\begin{equation}
  \mathcal{L}_1 = -\log \frac{\exp\big(\cos(\mathbf{z}_q, \mathbf{z}_{d^+}) / \tau_1\big)}{\sum_{d \in \mathcal{B}} \exp\big(\cos(\mathbf{z}_q, \mathbf{z}_{d}) / \tau_1\big)}
\end{equation}

where $\mathcal{B} \subset D_0$ is the set of documents in the batch, $\tau_1$ is a temperature, and $d^+$ is the relevant document(s) paired with $q$. For each query, the other documents in $\mathcal{B}$ serve as in-batch negatives. We denote the resulting encoder as $M_1$, which is used for DocID derivation for documents in both $D_0$ and $D_1$.

\subsubsection{Query-to-DocID optimization}

\paragraph{Seen documents} This stage is where our method departs most from prior work. Generative retrieval models (using either textual or numeric identifiers) typically perform two types of training: Indexing and retrieving. In prior work, the use of numeric identifiers requires the indexing training because the the embeddings of the identifier tokens are learned from scratch. Our DocID bit vectors are instead computed by the encoder from the document itself, so no indexing training is required. 
A query $q$ is encoded into its tanh output $\mathbf{z}_q$ by an encoder whose weights are initialized from $M_1$ (see above). We define the weighted query representation as
\begin{equation}\label{mu}
  \mathbf{u}_q= \mathbf{w} \odot \mathbf{z}_q
\end{equation}
where $\mathbf{z}_q \in (-1,1)^k$ is the encoder output for the query (using the same architecture as in equation \ref{archi}), $\odot$ is the element-wise product, $\mathbf{w} \in \mathbb{R}^k$ is a learnable vector with one entry per bit, shared across all queries, that weighs each bit's contribution to the ranking score. We denote this encoder together with $\mathbf{w}$ as the retriever encoder $M_2$. The relevance scores of a document $d$ for a given query $q$ is
\begin{equation}
  s(q,d) = \mathbf{b}_d^{\top} \mathbf{u}_q
\label{eq:rel_score}
\end{equation}

Stacking the bit vectors of all documents in $D_0$ as the rows of
$\mathbf{B} \in \{-1,+1\}^{|D_0| \times k}$ gives every score in a single matrix product. The retriever encoder $M_2$ is optimized with InfoNCE over documents $D_0$:

\begin{equation}
  \mathcal{L}_2 = -\log \frac{\exp\big(s(q,d^+) / \tau_2'(q)\big)}{\sum_{d \in D_0} \exp\big(s(q,d) / \tau_2'(q)\big)},
  \qquad
\end{equation}
where $\tau_2'(q) = \tau_2 \cdot  \big(\sqrt{k}\,
\|\mathbf{u}_q\|\big)$, with $\tau_2$ a temperature. $\sqrt{k}\,\|\mathbf{u}_q\|$ normalizes the dot product $s(q,d)$ and converts it to the cosine similarity between $\mathbf{b}_d$ and $\mathbf{w} \odot \mathbf{z}_q$. This loss trains each query to score its relevant DocID above all others in $D_0$. 

\paragraph{Unseen documents} The Query-to-DocID stage optimizes $M_2$ only on seen (training) documents $D_0$, as the unseen documents $D_1$ are introduced only at test time. This training setup may bias $M_2$ toward the seen documents, potentially limiting generalization to unseen documents.
To reduce the bias in the retrieval model, we adopt weight-space ensembling following WiSE-FT~\citep{wortsman2022robust}, which improves accuracy under distribution shift while preserving in-distribution accuracy. We interpolate the parameters of $M_1$ and $M_2$ to get our retriever $M_2'$: $\theta'_2 = \alpha\,\theta_1 + (1-\alpha)\,\theta_2$, where $\theta_1$ and $\theta_2$ are the parameters of $M_1$ and $M_2$, respectively. Since $M_1$ has no per-bit scale, we set its scale to $\mathbf{1} \in \mathbb{R}^{k}$. The merged encoder $M_2'$ is our retriever model and $M_2$ is discarded. We set $\alpha = 0.5$, following WiSE-FT, without tuning it on our data.

\subsubsection{Inference: multi-tree constrained beam search}

Both training stages operate on bits, but we perform inference token by token: each token represents an $m$-bit segment of $\mathbf{b}_d$ as an integer in $[0,2^m)$. Thus, each DocID has $T$ tokens and $k=Tm$ bits in total. This reduces the number of pruning decisions, as $m$ bits (instead of one bit) are used for each decision. Since no token score depends on previously decoded tokens, no autoregressive decoder is needed.

\paragraph{Token scoring}
In generative retrieval, a DocID is decoded as a sequence of tokens, each step producing a softmax over the possible token values, with a prefix tree restricting the output to valid DocIDs. Our add-on is to bit-factorize each token value into an $m$-bit pattern. 

Let the DocID of document $d$ be $c_1,\ldots,c_T$, with $c_t \in \{0,\ldots,2^m-1\}$. We map each token to its $m$-bit representation using a fixed table $\mathbf{P} \in \{-1,+1\}^{2^m \times m}$, whose row $\mathbf{p}_v$ is the binary code of $v$ in $\pm 1$ form. The bit-factorized DocID is then $\mathbf{p}_{c_1},\ldots,\mathbf{p}_{c_T}$.

Let $\mathbf{u}_t \in \mathbb{R}^m$ contain the entries of $\mathbf{u}_q$ (Equation~\ref{mu}) corresponding to token $t$, so that $\mathbf{u}_q = [\mathbf{u}_1;\ldots;\mathbf{u}_T]$. 
Recall that $s(q,d)$ is a sum over the $k$ bits (see Eq. \ref{eq:rel_score}). Since each token represents $m$ bits, this sum can be split into $T$ terms:
\begin{equation}
  s(q,d) = \sum_{t=1}^{T} \mathbf{p}_{c_t}^{\top} \mathbf{u}_t 
\end{equation}
The score is a sum of per-token terms, each depending only on a token value and the query.

At decoding time the document is not known in advance, so at step $t$ we score every
candidate value $v$ by $\mathbf{p}_v^{\top}\mathbf{u}_t$, and apply a softmax over the $2^m$ values:
\begin{equation}
  \log p_t(v \mid q) = \mathbf{p}_v\,\mathbf{u}_t - \log \sum_{v'=0}^{2^m-1} \exp\big(\mathbf{p}_{v'}\,\mathbf{u}_t\big),
\end{equation}

The second term is shared by all token values at level $t$ (denoted $R_t(q)$ for brevity),
so token values are ranked by $\mathbf{p}_v^{\top}\mathbf{u}_t$ alone. Summing along document $d$'s path gives:

\begin{equation}
\begin{aligned}
  \sum_{t=1}^{T} \log p_t(c_t \mid q) 
  &= \sum_{t=1}^{T} \mathbf{p}_{c_t}\,\mathbf{u}_t - R_t(q) 
  \\
  &= s(q,d) - R(q),
\end{aligned}
\end{equation}
where $R(q)$ is the same for every document, so the relative scores across documents are determined solely by $s(q,d)$. Normalization is optional because it does not affect the ranking.  In our main experiments, we use $T=10$ tokens of $m=12$ bits each, giving $k=120$ total bits.

\paragraph{Multi-trees beam search.} Beam search scores DocIDs token by token on the partial sum
$\sum_{t' \le t} \log p_{t'}(c_{t'} \mid q)$ and prunes prefixes as it goes. One potential problem is that pruning may discard a relevant DocID early, even if its remaining bits match the query well.  Because our tokens are independently scored through bit factorization, any bit permutation, followed by regrouping into tokens, generates a valid prefix tree. We therefore build $h$ trees using different bit permutations, run beam search on each tree, and rank the union of their candidates by log-probability. Since different bit orders prune based on different bits, using multiple trees can recover documents that a single tree would miss. We set $h = 8$ in our main experiments and analyze the effect of different values.


\section{Experimental Settings}




\subsection{Datasets}

We follow prior work \citep{tay2022transformer,wang2022neural,zhang2025replication, zhang2026model} to use Natural Questions (NQ320K) \citep{kwiatkowski2019natural}, MS MARCO \citep{nguyen2016ms}, and TriviaQA \citep{joshi2017triviaqa} in our experimetns. NQ320K and TriviaQA  include approximately 109K and 73K documents, respectively. To simulate newly arrived documents, we split the documents into seen ($D_0$) and unseen ($D_1$) sets, with (60$\%$, 40$\%$) in our main experiments and (90$\%$, 10$\%$) in additional experiments. For MS MARCO, we use the 320K-document and the (90$\%$, 10$\%$) split from \citet{zhang2026model}.

We evaluate by testing on queries whose gold documents are exclusively in either $D_0$ or $D_1$. Neither $D_1$ nor its associated queries are used during training or model selection. Further details on dataset sampling are provided in Appendix~\ref{data_splits}.

\subsection{Models}
We use T5-base (110M parameters) \citep{t5_model} and BERT-base-uncased (110M parameters) \citep{bert_model} as the encoders in all experiments, which are half the size as the T5-base encoder-decoder (220M) used by common baselines \citep{tay2022transformer,wang2022neural, sun2023learning}.

\subsection{Evaluation}
We follow prior work \citep{tay2022transformer,wang2022neural,sun2023learning} to report Hit$@$N, which is defined as the proportion of queries for which the relevant document appears in the top-N retrieved results. Because distinct documents can share one DocID, we solve such ties by expectation~\citep{mcsherry2008computing}: All documents sharing a retrieved identifier are returned, and each metric is its expected value over all orderings of the tied documents.


\begin{table*}[t]
\centering
\resizebox{0.85\textwidth}{!}{%
\begin{tabular}{llcccccccc}
\toprule
\multirow{2}{*}{Method} & \multirow{2}{*}{DocID type} & \multicolumn{4}{c}{NQ320K} & \multicolumn{4}{c}{TriviaQA} \\
\cmidrule(lr){3-6} \cmidrule(lr){7-10}
 & & Hit@1 & Hit@10 & Hit@100 & MRR@100 & Hit@1 & Hit@10 & Hit@100 & MRR@100 \\
\midrule
\rowcolor{gray!15}
\multicolumn{10}{l}{\textit{Queries with unseen target documents ($D_1$)}} \\
NCI & Hier. $k$-means & 0.07 & 0.61 & 1.97 & 0.24 & 0.13 & 0.71 & 2.48 & 0.36 \\
DSI & Hier. $k$-means & 0.07 & 0.61 & 3.51 & 0.28 & 0.09 & 0.73 & 2.80 & 0.33 \\
NCI (shared IDs) & Hier. $k$-means & 0.65 & 4.89 & 15.18 & 2.11 & 0.58 & 4.76 & 16.29 & 2.01 \\
DSI (shared IDs) & Hier. $k$-means & 0.58 & 4.20 & 11.24 & 1.80 & 0.43 & 3.69 & 12.21 & 1.55 \\
DSI & RQ-VAE  & 0.08 & 0.54 & 3.75 & 0.29 & 0.02 & 0.34 & 2.38 & 0.14 \\
ULTRON-PQ & Product quant. & 0.04 & 0.45 & 13.98 & 0.37 & 0.00 & 0.03 & 8.64 & 0.12 \\
MEVI & RQ & 0.66 & 5.18 & 16.06 & 2.17 & 0.67 & 5.04 & 16.09 & 2.15 \\
ULTRON-URL & URL / title string & 13.12 & 27.41 & 43.36 & 18.18 & -- & -- & -- & -- \\
SEAL (BART) & Substrings & 10.76 & 50.32 & 78.93 & 23.72 & 26.77 & 59.03 & 79.75 & 37.67 \\
CoBIGR-Bert & Bit-factorized code & \underline{41.45}\rlap{$^{*}$} & \textbf{77.74}\rlap{$^{*}$} & \textbf{85.48}\rlap{$^{*}$} & \textbf{54.79}\rlap{$^{*}$} & \textbf{53.13}\rlap{$^{*}$} & \textbf{77.15}\rlap{$^{*}$} & \textbf{85.22}\rlap{$^{*}$} & \textbf{61.88}\rlap{$^{*}$} \\
CoBIGR-T5 & Bit-factorized code & \textbf{42.90}\rlap{$^{*}$} & \underline{74.92}\rlap{$^{*}$} & \underline{81.80}\rlap{$^{*}$} & \underline{54.54}\rlap{$^{*}$} & \underline{49.72}\rlap{$^{*}$} & \underline{74.74}\rlap{$^{*}$} & \underline{82.93}\rlap{$^{*}$} & \underline{58.66}\rlap{$^{*}$} \\
\cmidrule{1-10}
\multicolumn{10}{l}{\textit{Reference (non-GenIR approach)}} \\
BM25 & Term weights (sparse) & 36.20 & 69.73 & 87.68 & 47.55 & 47.75 & 77.48 & 89.86 & 58.53 \\
GTR & Dense vector & 54.39 & 83.84 & 93.65 & 64.86 & 56.77 & 80.12 & 91.02 & 65.32 \\
\midrule
\rowcolor{gray!15}
\multicolumn{10}{l}{\textit{Queries with seen target documents ($D_0$)}} \\
NCI & Hier. $k$-means & 68.83 & \underline{87.39} & \textbf{93.75} & 75.69 & 68.82 & \textbf{89.51} & \textbf{95.27} & 76.69 \\
DSI & Hier. $k$-means & 63.81 & 79.15 & 85.65 & 69.62 & 58.90 & 78.37 & 86.63 & 66.20 \\
DSI & RQ-VAE  & 63.26 & 80.71 & 86.97 & 69.92 & 60.53 & 79.73 & 87.76 & 67.54 \\
ULTRON-PQ & Product quant. & 64.92 & 81.74 & 87.84 & 71.38 & 68.71 & 84.35 & 89.86 & 74.73 \\
MEVI & RQ & 50.86 & 80.53 & 88.93 & 61.98 & 56.98 & 83.58 & 90.80 & 66.99 \\
ULTRON-URL & URL / title string & 66.11 & 81.04 & 88.01 & 71.81 & -- & -- & -- & -- \\
SEAL (BART) & Substrings & 67.73 & 84.28 & \underline{92.61} & 74.01 & \textbf{73.18} & 85.84 & \underline{91.90} & \textbf{77.93} \\
CoBIGR-Bert & Bit-factorized code & \textbf{71.06}\rlap{$^{*}$} & \textbf{87.59} & 91.87 & \textbf{77.18}\rlap{$^{*}$} & \underline{70.73} & \underline{86.76} & 91.18 & \underline{76.75} \\
CoBIGR-T5 & Bit-factorized code & \underline{69.68} & 85.80 & 89.77 & 75.68 & 70.05 & 85.73 & 90.10 & 76.04 \\
\cmidrule{1-10}
\multicolumn{10}{l}{\textit{Reference (non-GenIR approach)}} \\
BM25 & Term weights (sparse) & 36.49 & 69.13 & 87.62 & 47.68 & 49.18 & 79.15 & 90.97 & 60.00 \\
GTR & Dense vector & 54.26 & 83.84 & 93.79 & 64.92 & 57.84 & 81.61 & 91.73 & 66.60 \\
\bottomrule
\end{tabular}
}
\caption{Retrieval results on NQ320K and TriviaQA with $|D_0|{=}60\%$ seen / $|D_1|{=}40\%$ unseen split. MEVI: we reproduce only the generative-stage results without DPR ranking. ULTRON-URL: TriviaQA lacks URLs. $^{*}$: ours are significantly better than both NCI and SEAL (paired $t$-test, $p<0.05$ with Bonferroni correction). Reference retrievers search the full corpus with larger indexes: 3 KB/doc for GTR and 1.4 KB/doc for BM25, vs. 0.015 KB/doc for ours.}

\label{tab:dyn40v2}
\end{table*}

\subsection{Baselines}

\update{All baselines are generative retrievers, the paradigm we follow. Our core research question is whether an alternative numeric DocID design improves generalization to newly added documents, so we focus on numeric DocID baselines.}

Following recent studies of generative retrieval over dynamic corpora~\citep{zhang2025replication, zhang2026model}, we cover four numeric DocID designs:  \textbf{hierarchical $k$-means}, used by DSI~\citep{tay2022transformer} and NCI~\citep{wang2022neural}; \textbf{PQ}, used by ULTRON-PQ~\citep{zhou2022ultron}; \textbf{RQ}, used by MEVI~\citep{zhang2023mevi};  \textbf{RQ-VAE}, used by  TIGER~\citep{rajput2023recommender}. For further comparison, we include textual DocID baselines, ULTRON-URL~\citep{zhou2022ultron} and SEAL~\citep{bevilacqua2022autoregressive}, and weight-update methods DSI++~\citep{mehta2023dsi++} and DOME~\citep{zhang2026model}. This allows us to compare our zero-shot adaptation approach with methods that update the weights using $D_1$. See Appendix~\ref{baselines} for details. Non-generative approaches baselines such as BM25 \citep{robertson2009probabilistic} and GTR \citep{ni2022large} are reported as references, since the different indexing and exhaustive search limit direct comparison.

We reproduce all baselines on NQ320K and TriviaQA, while using results reported in prior work \citep{zhang2026model} for MS MARCO.

\subsection{Hyperparameters}

In our setting, queries and documents are truncated to 128 tokens for DocID derivation, with an effective batch size of 512. We train using pseudo-queries with a budget of 96 A100 GPU-hours. Other hyperparameters are detailed in Appendix~\ref{hyper_appen}. For other baselines, we follow their original settings and provide the same pseudo-queries and GPU budget.

During inference, following prior work~\citep{wang2022neural}, we use a beam size of 100 for reproduced baselines, except SEAL, which uses its default size of 15~\citep{bevilacqua2022autoregressive} because each token can already map to many documents.

\section{Results}

\paragraph{Comparison across identifier types}

Table~\ref{tab:dyn40v2} reports the results for two query groups under the $60\%/40\%$ split on the NQ320K and TriviaQA datasets: queries whose gold documents are in $D_1$ (unseen document, upper part) and those in $D_0$ (seen document, lower part). We first focus on $D_1$ queries, which directly evaluate generalization to unseen documents. Our method achieves the highest performance among the compared DocID designs on both datasets by a large margin. Hierarchical $k$-means DocIDs generalize poorly because their final identifier digit acts as a document-specific counter, making new identifiers difficult to predict. Removing the counter digit improves both methods slightly (indicated with `shared IDs' in parentheses), but their performance remains limited. Quantization-based DocIDs also show poor generalization to unseen documents. Our DocID design also substantially improves over the text-based DocID methods ULTRON-URL and SEAL for unseen documents.

For queries with gold documents in $D_0$, our method achieves higher Hit@1 and MRR@100, indicating stronger top-ranking performance, while NCI achieves higher Hit@100, suggesting better high-rank retrieval coverage. The $90\%/10\%$ splits shows the similar patterns (Appendix \ref{dyn10v2}). We also report results on the full corpus in Appendix~\ref{full_corpus}. Overall, our method maintains comparable performance to baselines on seen documents while improving with a large margin on unseen documents.

\paragraph{Comparison with incremental training and model editing}
Table \ref{tab:msmarco9010} shows that our method maintains similar performance on $D_0$. More importantly, our zero-shot adaptation achieves results within approximately 3 percentage points of the incremental training and model editing methods on $D_1$. This suggests that our method can reduce the need for additional training while maintaining competitive performance.

\begin{table}[t]
\centering
\resizebox{0.25\textwidth}{!}{%
\begin{tabular}{lcc}
\toprule
Method & Hit@10 & MRR@10 \\
\midrule
\rowcolor{gray!15}
\multicolumn{3}{l}{\textit{Queries with target documents in $D_1$}} \\
DSI++$^\dagger$ & 75.9 & 51.7 \\
DOME$^\dagger$ & \textbf{76.4} & \textbf{52.4} \\
CoBIGR-Bert & 74.5 & 49.1 \\
CoBIGR-T5 & 74.7 & 48.0 \\
\midrule
From Scratch & 77.2 & 52.9 \\
\midrule
\rowcolor{gray!15}
\multicolumn{3}{l}{\textit{Queries with target documents in $D_0$}} \\
DSI++ & 77.5 & 53.2 \\
DOME & 77.5 & 53.2 \\
CoBIGR-Bert & \textbf{78.2} & \textbf{56.2} \\
CoBIGR-T5 & 78.1 & 55.6 \\
\midrule
From Scratch & 77.5 & 53.2 \\
\bottomrule
\end{tabular}
}
\caption{Comparison with weight-update methods on MS MARCO, with $|D_0|{=}90\%$  / $|D_1|{=}10\%$. Runs marked with $\dagger$ are not a direct comparison with CoBIGR, yet shown as a reference; they use $D_1$ to update model weights, while our method is zero-shot adaptation on unseen corpus $D_1$.  All baselines are cited from \citet{zhang2026model}. From Scratch is the paper's upper bound with full-parameter updates.}
\label{tab:msmarco9010}
\end{table}

\paragraph{Efficiency} \update{Our method provides efficient retrieval: token scores for all positions come from one encoder forward pass, so decoding needs no autoregressive decoder calls.} We measure single-query latency (batch size 1) on a single A100 GPU on the NQ320K dataset, and find that our method takes 24 ms per query and is at least $7\times$ faster than NCI, DSI, and MEVI, $28\times$ faster than ULTRON, and $397\times$ faster than SEAL. See Appendix~\ref{pareto_hit1} for the overall efficiency analysis. On MS MARCO, whose corpus is about three times larger, latency increases by only 4 ms per query.

\section{Ablation}

\paragraph{Effect of weight interpolation}

Table \ref{tab:ablation_stages} shows that the DocID encoder $M_1$ provides better generalization to unseen documents, while the retriever $M_2$, which is optimized on seen documents, achieves better performance on seen documents, with Hit@1 increasing by more than 10 percentage points on both datasets. However, $M_2$ also shows a substantial drop in performance on unseen documents. 
Our final $M_2'$ model provides a balance between the two. Additionally, using multiple trees substantially improves almotst all metrics. This suggests 
permutation helps reduce premature pruning of relevant documents.

\begin{table*}[t]
\centering
\resizebox{1\textwidth}{!}{%
\begin{tabular}{lcccccccc}
\toprule
\multirow{2}{*}{Model} & \multicolumn{4}{c}{NQ320K} & \multicolumn{4}{c}{TriviaQA} \\

\cmidrule(lr){2-5} \cmidrule(lr){6-9}
 & Hit@1 & Hit@10 & Hit@100 & MRR@100 & Hit@1 & Hit@10 & Hit@100 & MRR@100 \\
\midrule
$M_1$ (DocID encoder) & 57.03 / \textbf{44.12} & 73.69 / 63.31 & 74.50 / 64.46 & 63.67 / 51.41 & 51.69 / 43.95 & 65.27 / 56.95 & 66.02 / 58.19 & 57.03 / 48.86 \\
$M_2$ (trained on $D_0$) & 68.58 / 20.71 & 76.60 / 44.81 & 76.89 / 45.49 & 71.75 / 30.42 & 64.54 / 36.51 & 71.74 / 49.62 & 72.15 / 50.61 & 67.49 / 41.62 \\
$M_2'$ (weight interpolation) & 67.69 / 36.09 & 78.20 / 58.83 & 78.78 / 59.77 & 71.93 / 45.10 & 62.20 / 42.86 & 71.34 / 55.97 & 71.90 / 57.08 & 65.90 / 47.87 \\
\midrule
$M_2'$ + multi-tree ($h{=}8$) & \textbf{71.06} / 41.45 & \textbf{87.59} / \textbf{77.74} & \textbf{91.87} / \textbf{85.48} & \textbf{77.18} / \textbf{54.79} & \textbf{70.73} / \textbf{53.13} & \textbf{86.76} / \textbf{77.15} & \textbf{91.18} / \textbf{85.22} & \textbf{76.75} / \textbf{61.88} \\
\bottomrule
\end{tabular}%
}
\caption{Ablation on NQ320K and TriviaQA ($|D_0|{=}60\%$, $|D_1|{=}40\%$) with the BERT backbone. Identifiers are
derived by $M_1$ in every row. All rows retrieve by constrained beam search (beam 100); the last row runs the same search over $h{=}8$ permuted trees.}
\label{tab:ablation_stages}
\end{table*}

\begin{figure}[t]
        \centering
        \includegraphics[width=0.5\textwidth]{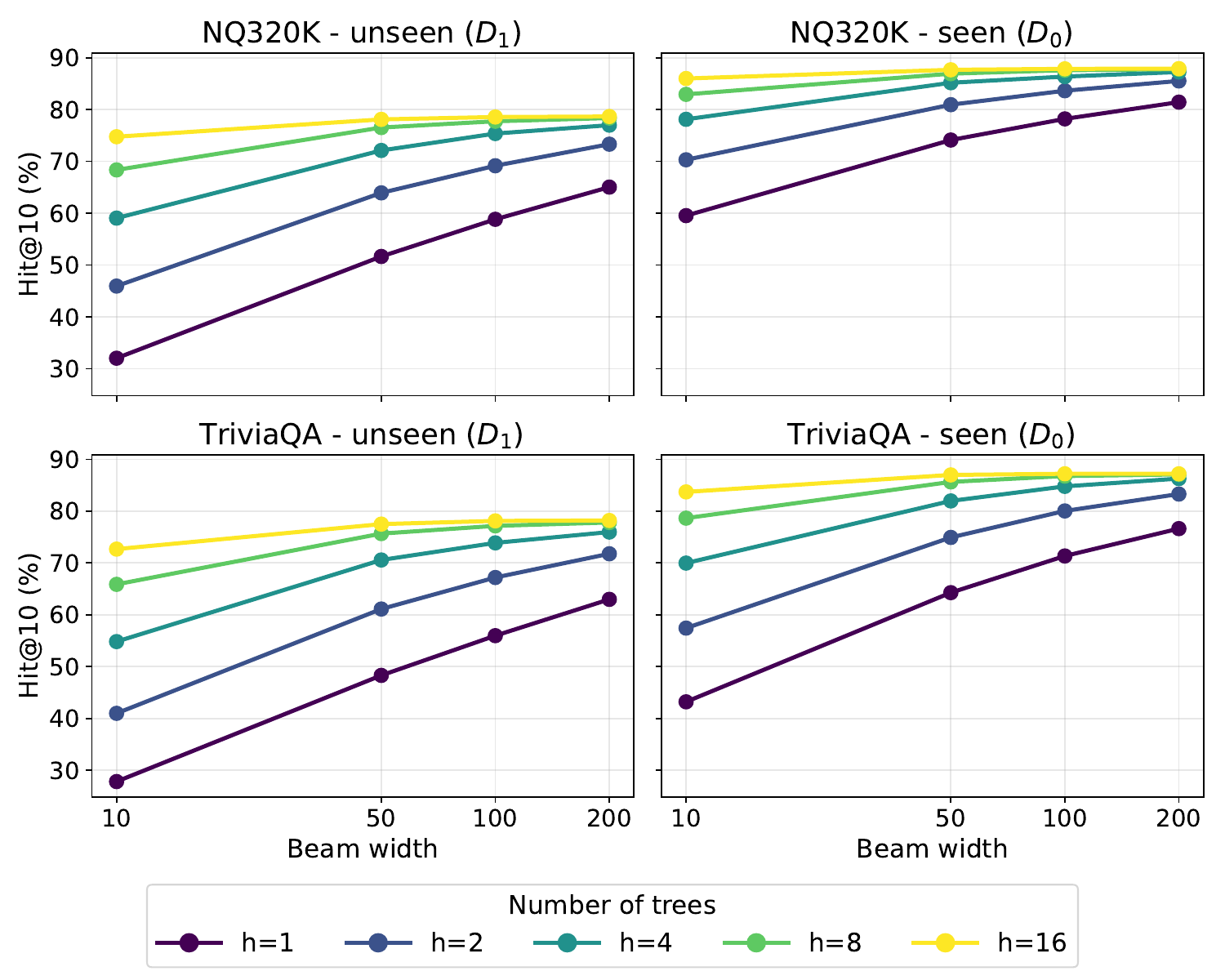}
        \caption{Hit@10 by different size of trees and beams ($|D_0|{=}60\%$, $|D_1|{=}40\%$) with the BERT backbone.}
        \label{fig:R10_beam_tree}
\end{figure}

\paragraph{Effect of multiple trees} 
Figures \ref{fig:R10_beam_tree} shows Hit@10 under different beam size and permutation-tree counts. Increasing both the beam size and the number of trees clearly improves the metric. In particular, increasing the number of trees brings larger improvements than increasing the beam size. Across both metrics and datasets, increasing the beam size from 10 to 200 (20$\times$) with the tree size fixed at $h=1$ leads to smaller improvements than increasing the tree size from $h=1$ to 16 (16$\times$) with the beam size fixed at 10. See Appendix \ref{m_tree} for other metrics .

\begin{figure}[t]
        \centering
        \includegraphics[width=0.5\textwidth]{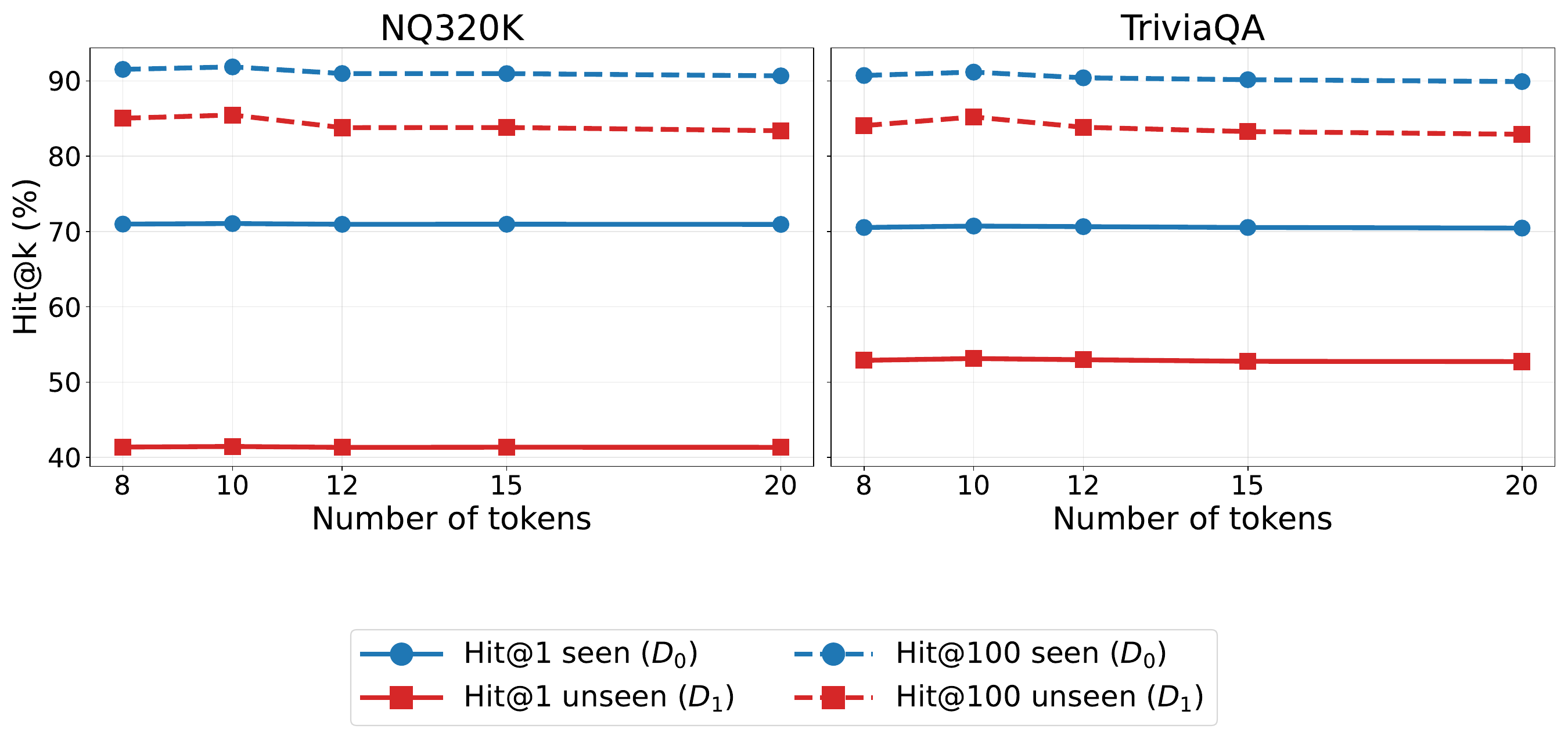}
        \caption{Results with different bits per token, given 120 bits ($|D_0|{=}60\%$, $|D_1|{=}40\%$) with the BERT backbone.}
        \label{fig:bits_per_token}
\end{figure}

\begin{figure}[t]
        \centering
        \includegraphics[width=0.35\textwidth]{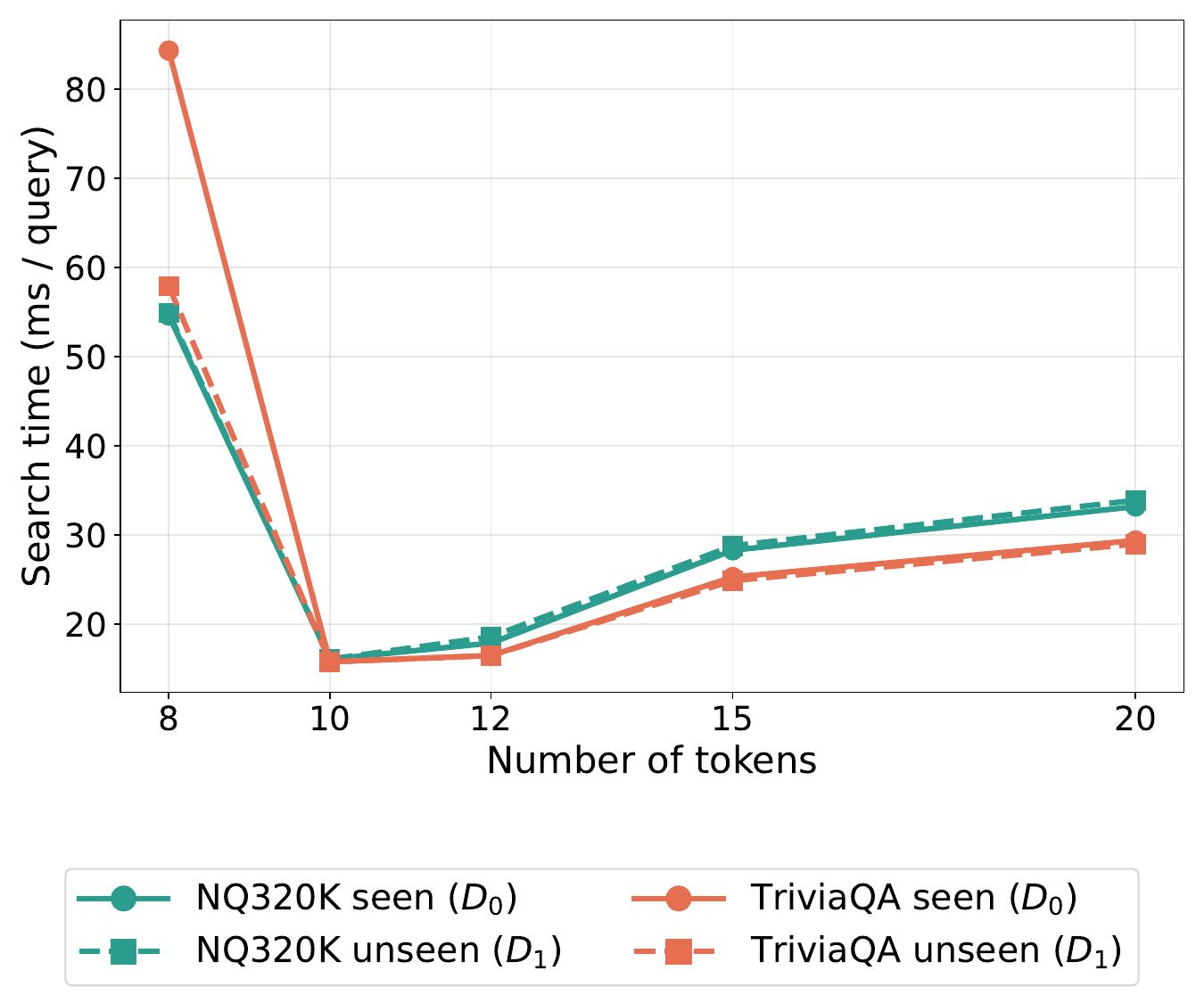}
        \caption{Search latency by tokens $T$ given 120 bits ($|D_0|{=}60\%$, $|D_1|{=}40\%$) with the BERT backbone. Query encoding time is excluded, as it is the same for all settings.}
        \label{fig:bits_per_token_ms}
\end{figure}

\paragraph{Effect of token length}\label{token_length}
In the main experiments, each token represents 12 bits, resulting in a total of 10 tokens ($=\frac{120}{12}$). Since the score is a sum over bits, the same identifiers can be read as $T$ tokens of $m$ bits for any $m$ dividing 120, without retraining; we additionally evaluate $m \in \{6, 8, 10, 15\}$.  Figure \ref{fig:bits_per_token} shows that different grouping sizes achieve similar results. However, the number of bits per token has a clear effect on inference speed: Figure \ref{fig:bits_per_token_ms} shows using too few or too many bits per token can reduce decoding speed. Nevertheless, search remains fast overall.

\begin{figure}[t]
        \centering
        \includegraphics[width=0.5\textwidth]{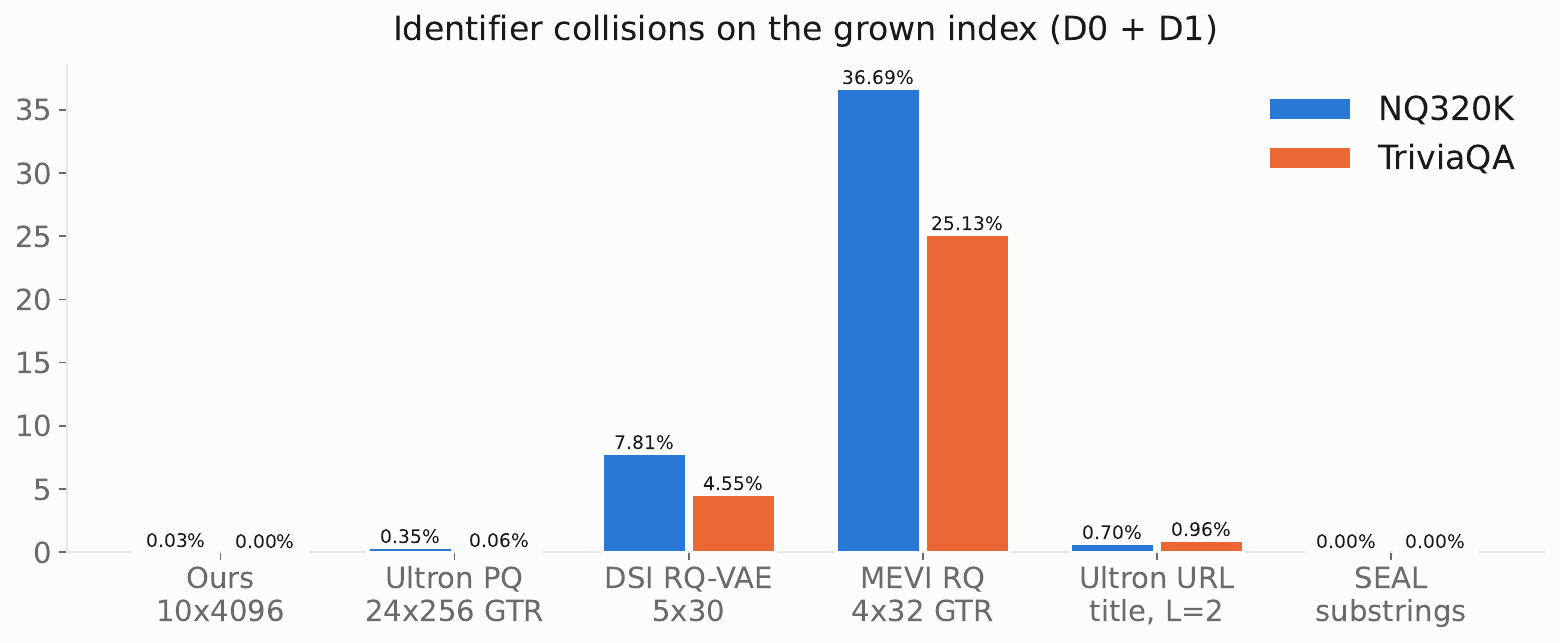}
        \caption{Number of document collisions ($|D_0|{=}60\%$, $|D_1|{=}40\%$) with the BERT backbone.}
        \label{fig:collision_doc}
\end{figure}

\paragraph{Document collision} \label{collision}
Figure \ref{fig:collision_doc} shows the number of documents sharing the same DocID on NQ320K and TriviaQA. All baselines have collision rates below 10$\%$, while our method achieves the lowest collision rate among the numeric ID baselines (32 documents for NQ and 1 documents for TriviaQA). SEAL has no duplicate identifiers because it assigns no code at all. A manual inspection shows that (1) no more than two documents share a DocID, and (2) Most collisions occur between near duplicate documents, indicating good separation of dissimilar documents.

\section{Conclusions}

We propose a computable DocID design whose bit-level factorization allows new DocIDs to be scored, while model fusion and multiple-tree search enable better zero-shot retrieval.
We find that existing numeric DocID designs perform well on seen documents but degrade substantially on unseen documents, while our approach maintains comparable performance on seen documents and improves on unseen documents by a large margin over previous methods, and remains close to methods that update model weights.
This suggests that our DocID design for incorporating new documents without retraining is promising, especially given the high computational cost of retraining generative retrievers.
In future work, we plan to apply factorization to hierarchical DocIDs (RQ-VAE or H-KMEANS) to investigate whether it can further improve recall at high ranks.

\clearpage

\section*{Acknowledgments}
This publication is part of the project LESSEN with project number NWA.1389.20.183 of the research program NWA ORC 2020/21 which is (partly) financed by the Dutch Research Council (NWO).

\section*{Limitations}

\textbf{Flat identifiers make early pruning harder.} In our token design, no prefix is more informative than another, so early decoding prunes on limited information. Multi-tree beam search reduces this effect, but Hit@100 on seen documents remains slightly below some baselines. In future work, we plan to combine bit factorization with residual quantization such as RQ-VAE, whose coarse-to-fine codes may make early levels more informative and reduce pruning errors.

\textbf{Language.} Like most of the prior work, we only focus on English datasets. This is common because most of benchmark datasets are in English. Our method is completely data-driven and not dependent on labeled data and it is therefore applicable to other languages than English.\\

\noindent

\noindent

\bibliography{anthology,custom}

@inproceedings{kim2024exploring,
  title={Exploring the practicality of generative retrieval on dynamic corpora},
  author={Kim, Chaeeun and Yoon, Soyoung and Lee, Hyunji and Jang, Joel and Yang, Sohee and Seo, Minjoon},
  booktitle={Proceedings of the 2024 Conference on Empirical Methods in Natural Language Processing},
  pages={13616--13633},
  year={2024}
}

@article{li2025matching,
  title={From matching to generation: A survey on generative information retrieval},
  author={Li, Xiaoxi and Jin, Jiajie and Zhou, Yujia and Zhang, Yuyao and Zhang, Peitian and Zhu, Yutao and Dou, Zhicheng},
  journal={ACM Transactions on Information Systems},
  volume={43},
  number={3},
  pages={1--62},
  year={2025},
  publisher={ACM New York, NY}
}

@inproceedings{zhang2025replication,
  title={Replication and exploration of generative retrieval over dynamic corpora},
  author={Zhang, Zhen and Ma, Xinyu and Sun, Weiwei and Ren, Pengjie and Chen, Zhumin and Wang, Shuaiqiang and Yin, Dawei and de Rijke, Maarten and Ren, Zhaochun},
  booktitle={Proceedings of the 48th international ACM SIGIR conference on research and development in information retrieval},
  pages={3325--3334},
  year={2025}
}

@article{tay2022transformer,
  title={Transformer memory as a differentiable search index},
  author={Tay, Yi and Tran, Vinh and Dehghani, Mostafa and Ni, Jianmo and Bahri, Dara and Mehta, Harsh and Qin, Zhen and Hui, Kai and Zhao, Zhe and Gupta, Jai and others},
  journal={Advances in neural information processing systems},
  volume={35},
  pages={21831--21843},
  year={2022}
}

@article{wang2022neural,
  title={A neural corpus indexer for document retrieval},
  author={Wang, Yujing and Hou, Yingyan and Wang, Haonan and Miao, Ziming and Wu, Shibin and Chen, Qi and Xia, Yuqing and Chi, Chengmin and Zhao, Guoshuai and Liu, Zheng and others},
  journal={Advances in Neural Information Processing Systems},
  volume={35},
  pages={25600--25614},
  year={2022}
}

@article{rajput2023recommender,
  title={Recommender systems with generative retrieval},
  author={Rajput, Shashank and Mehta, Nikhil and Singh, Anima and Hulikal Keshavan, Raghunandan and Vu, Trung and Heldt, Lukasz and Hong, Lichan and Tay, Yi and Tran, Vinh and Samost, Jonah and others},
  journal={Advances in Neural Information Processing Systems},
  volume={36},
  pages={10299--10315},
  year={2023}
}

@article{sun2023learning,
  title={Learning to tokenize for generative retrieval},
  author={Sun, Weiwei and Yan, Lingyong and Chen, Zheng and Wang, Shuaiqiang and Zhu, Haichao and Ren, Pengjie and Chen, Zhumin and Yin, Dawei and Rijke, Maarten and Ren, Zhaochun},
  journal={Advances in Neural Information Processing Systems},
  volume={36},
  pages={46345--46361},
  year={2023}
}

@article{zhang2023mevi,
  title={Model-enhanced vector index},
  author={Zhang, Hailin and Wang, Yujing and Chen, Qi and Chang, Ruiheng and Zhang, Ting and Miao, Ziming and Hou, Yingyan and Ding, Yang and Miao, Xupeng and Wang, Haonan and others},
  journal={Advances in Neural Information Processing Systems},
  volume={36},
  pages={54903--54917},
  year={2023}
}

@article{zhao2025unifying,
  title={Unifying Search and Recommendation with Dual-View Representation Learning in a Generative Paradigm},
  author={Zhao, Jujia and Wang, Wenjie and Xu, Chen and Chen, Xiuying and Ren, Zhaochun and Verberne, Suzan},
  journal={ACM Transactions on Information Systems},
  year={2025},
  publisher={ACM New York, NY}
}

@article{de2020autoregressive,
  title={Autoregressive entity retrieval},
  author={De Cao, Nicola and Izacard, Gautier and Riedel, Sebastian and Petroni, Fabio},
  journal={ICLR 2021},
  year={2020}
}

@article{bevilacqua2022autoregressive,
  title={Autoregressive search engines: Generating substrings as document identifiers},
  author={Bevilacqua, Michele and Ottaviano, Giuseppe and Lewis, Patrick and Yih, Scott and Riedel, Sebastian and Petroni, Fabio},
  journal={Advances in Neural Information Processing Systems},
  volume={35},
  pages={31668--31683},
  year={2022}
}

@article{zhou2022ultron,
  title={Ultron: An ultimate retriever on corpus with a model-based indexer},
  author={Zhou, Yujia and Yao, Jing and Dou, Zhicheng and Wu, Ledell and Zhang, Peitian and Wen, Ji-Rong},
  journal={arXiv preprint arXiv:2208.09257},
  year={2022}
}

@inproceedings{wang2023novo,
  title={Novo: Learnable and interpretable document identifiers for model-based ir},
  author={Wang, Zihan and Zhou, Yujia and Tu, Yiteng and Dou, Zhicheng},
  booktitle={Proceedings of the 32nd ACM International Conference on Information and Knowledge Management},
  pages={2656--2665},
  year={2023}
}

@inproceedings{zhang2024generative,
  title={Generative retrieval via term set generation},
  author={Zhang, Peitian and Liu, Zheng and Zhou, Yujia and Dou, Zhicheng and Liu, Fangchao and Cao, Zhao},
  booktitle={Proceedings of the 47th international ACM SIGIR conference on research and development in information retrieval},
  pages={458--468},
  year={2024}
}

@inproceedings{askari2026generative,
  title={Generative retrieval with few-shot indexing},
  author={Askari, Arian and Meng, Chuan and Aliannejadi, Mohammad and Ren, Zhaochun and Kanoulas, Evangelos and Verberne, Suzan},
  booktitle={European Conference on Information Retrieval},
  pages={606--614},
  year={2026},
  organization={Springer}
}

@inproceedings{yamada2021efficient,
  title={Efficient passage retrieval with hashing for open-domain question answering},
  author={Yamada, Ikuya and Asai, Akari and Hajishirzi, Hannaneh},
  booktitle={Proceedings of the 59th Annual Meeting of the Association for Computational Linguistics and the 11th International Joint Conference on Natural Language Processing (Volume 2: Short Papers)},
  pages={979--986},
  year={2021}
}

@article{he2025survey,
  title={A Survey on Deep Text Hashing: Efficient Semantic Text Retrieval with Binary Representation},
  author={He, Liyang and Huang, Zhenya and Yang, Cheng and Li, Rui and Zhang, Zheng and Zhang, Kai and Li, Zhi and Liu, Qi and Chen, Enhong},
  journal={arXiv preprint arXiv:2510.27232},
  year={2025}
}

@inproceedings{zhang2026flash,
  title={FLASH: Fast Generative Retrieval via Autoregressive Semantic Hashing with Provably Distance Bounds},
  author={Zhang, Yifei and Zhu, Hao and Shi, Haoran and Chen, Yanyu and Han, Xiaolin and Zhang, Yupei and Song, Lingyun and Wang, Wenxuan and Zhou, Chao and Shang, Xuequn and others},
  booktitle={Proceedings of the 32nd ACM SIGKDD Conference on Knowledge Discovery and Data Mining V. 2},
  pages={6594--6605},
  year={2026}
}

@inproceedings{mehta2023dsi++,
  title={Dsi++: Updating transformer memory with new documents},
  author={Mehta, Sanket Vaibhav and Gupta, Jai and Tay, Yi and Dehghani, Mostafa and Tran, Vinh Q and Rao, Jinfeng and Najork, Marc and Strubell, Emma and Metzler, Donald},
  booktitle={Proceedings of the 2023 conference on empirical methods in natural language processing},
  pages={8198--8213},
  year={2023}
}

@inproceedings{kishore2023incdsi,
  title={Incdsi: Incrementally updatable document retrieval},
  author={Kishore, Varsha and Wan, Chao and Lovelace, Justin and Artzi, Yoav and Weinberger, Kilian Q},
  booktitle={International conference on machine learning},
  pages={17122--17134},
  year={2023},
  organization={PMLR}
}

@inproceedings{zhang2026model,
  title={Model Editing for New Document Integration in Generative Information Retrieval},
  author={Zhang, Zhen and Wang, Zihan and Ma, Xinyu and Wang, Shuaiqiang and Yin, Dawei and Xin, Xin and Ren, Pengjie and de Rijke, Maarten and Ren, Zhaochun},
  booktitle={Proceedings of the ACM Web Conference 2026},
  pages={1993--2003},
  year={2026}
}

@article{nguyen2016ms,
  title={Ms marco: A human-generated machine reading comprehension dataset},
  author={Nguyen, Tri and Rosenberg, Mir and Song, Xia and Gao, Jianfeng and Tiwary, Saurabh and Majumder, Rangan and Deng, Li},
  year={2016}
}

@inproceedings{joshi2017triviaqa,
  title={Triviaqa: A large scale distantly supervised challenge dataset for reading comprehension},
  author={Joshi, Mandar and Choi, Eunsol and Weld, Daniel S and Zettlemoyer, Luke},
  booktitle={Proceedings of the 55th Annual Meeting of the Association for Computational Linguistics (Volume 1: Long Papers)},
  pages={1601--1611},
  year={2017}
}

@article{kwiatkowski2019natural,
  title={Natural questions: a benchmark for question answering research},
  author={Kwiatkowski, Tom and Palomaki, Jennimaria and Redfield, Olivia and Collins, Michael and Parikh, Ankur and Alberti, Chris and Epstein, Danielle and Polosukhin, Illia and Devlin, Jacob and Lee, Kenton and others},
  journal={Transactions of the Association for Computational Linguistics},
  volume={7},
  pages={453--466},
  year={2019},
  publisher={MIT Press One Rogers Street, Cambridge, MA 02142-1209, USA journals-info~…}
}

@inproceedings{wortsman2022robust,
  title={Robust fine-tuning of zero-shot models},
  author={Wortsman, Mitchell and Ilharco, Gabriel and Kim, Jong Wook and Li, Mike and Kornblith, Simon and Roelofs, Rebecca and Lopes, Raphael Gontijo and Hajishirzi, Hannaneh and Farhadi, Ali and Namkoong, Hongseok and others},
  booktitle={2022 IEEE/CVF Conference on Computer Vision and Pattern Recognition (CVPR)},
  pages={7949--7961},
  year={2022},
  organization={IEEE}
}

@article{ba2016layer,
  title={Layer normalization},
  author={Ba, Jimmy Lei and Kiros, Jamie Ryan and Hinton, Geoffrey E},
  journal={arXiv preprint arXiv:1607.06450},
  year={2016}
}

@article{oord2018representation,
  title={Representation learning with contrastive predictive coding},
  author={Oord, Aaron van den and Li, Yazhe and Vinyals, Oriol},
  journal={arXiv preprint arXiv:1807.03748},
  year={2018}
}

@inproceedings{karpukhin2020dense,
  title={Dense passage retrieval for open-domain question answering},
  author={Karpukhin, Vladimir and Oguz, Barlas and Min, Sewon and Lewis, Patrick and Wu, Ledell and Edunov, Sergey and Chen, Danqi and Yih, Wen-tau},
  booktitle={Proceedings of the 2020 conference on empirical methods in natural language processing (EMNLP)},
  pages={6769--6781},
  year={2020}
}

@article{bert_model,
  author    = {Jacob Devlin and
               Ming{-}Wei Chang and
               Kenton Lee and
               Kristina Toutanova},
  title     = {{BERT:} Pre-training of Deep Bidirectional Transformers for Language
               Understanding},
  journal   = {CoRR},
  volume    = {abs/1810.04805},
  year      = {2018},
  url       = {http://arxiv.org/abs/1810.04805},
  archivePrefix = {arXiv},
  eprint    = {1810.04805},
  bibsource = {dblp computer science bibliography, https://dblp.org}
}

@article{t5_model,
  author  = {Colin Raffel and Noam Shazeer and Adam Roberts and Katherine Lee and Sharan Narang and Michael Matena and Yanqi Zhou and Wei Li and Peter J. Liu},
  title   = {Exploring the Limits of Transfer Learning with a Unified Text-to-Text Transformer},
  journal = {Journal of Machine Learning Research},
  year    = {2020},
  volume  = {21},
  number  = {140},
  pages   = {1-67},
  url     = {http://jmlr.org/papers/v21/20-074.html}
}

@article{nogueira2019doc2query,
  title={From doc2query to docTTTTTquery},
  author={Nogueira, Rodrigo and Lin, Jimmy and Epistemic, AI}
}

@inproceedings{mcsherry2008computing,
  title={Computing information retrieval performance measures efficiently in the presence of tied scores},
  author={McSherry, Frank and Najork, Marc},
  booktitle={European conference on information retrieval},
  pages={414--421},
  year={2008},
  organization={Springer}
}

@article{robertson2009probabilistic,
  title={The probabilistic relevance framework: BM25 and beyond},
  author={Robertson, Stephen and Zaragoza, Hugo},
  journal={Foundations and trends{\textregistered} in information retrieval},
  volume={4},
  number={1-2},
  pages={1--174},
  year={2009},
  publisher={Emerald Publishing Limited}
}

@inproceedings{ni2022large,
  title={Large dual encoders are generalizable retrievers},
  author={Ni, Jianmo and Qu, Chen and Lu, Jing and Dai, Zhuyun and Abrego, Gustavo Hernandez and Ma, Ji and Zhao, Vincent and Luan, Yi and Hall, Keith and Chang, Ming-Wei and others},
  booktitle={Proceedings of the 2022 Conference on Empirical Methods in Natural Language Processing},
  pages={9844--9855},
  year={2022}
}
\bibliographystyle{acl_natbib}
\clearpage

\appendix
\section{Appendix}
\subsection{Dataset statistics} \label{data_splits}

For NQ320K and TriviaQA, with each split, we distinguish seen document queries, whose gold documents are entirely in $D_0$, from unseen document queries, whose gold documents are entirely in $D_1$. Queries with gold documents in both sets are discarded. We randomly hold out 10,000 seen document queries for testing and 10 $\%$ of the remaining queries for validation, using the rest for training. We sample 10,000 unseen document queries for testing.\footnote{TriviaQA has only 4,195 eligible unseen queries at $|D_1|=10\%$, so we use all of them.} For MS MARCO, we follow the splits in  \citet{zhang2026model}. See Table \ref{tab:data_statistics} for details.

\begin{table*}[t]
\centering
\small
\caption{Statistics of the dynamic-corpus splits. Train and valid have all gold documents only in $D_0$. Test queries for $D_0$ are held out from training.}
\label{tab:split_stats}
\begin{tabular}{llrrrrrr}
\toprule
 & & \multicolumn{2}{c}{\#Documents} & \multicolumn{4}{c}{\#Queries} \\
\cmidrule(lr){3-4} \cmidrule(lr){5-8}
Dataset & $D_0$/$D_1$ & $|D_0|$ & $|D_1|$ & Train & Valid & Test (for $D_0$) & Test (for $D_1$) \\
\midrule
\multirow{2}{*}{NQ320K}   & 90/10 & 98,765 & 10,974 & 246,751 & 27,417 & 10,000 & 10,000 \\
                          & 60/40 & 65,843 & 43,896 & 162,077 & 18,009 & 10,000 & 10,000 \\
\midrule
\multirow{2}{*}{TriviaQA} & 90/10 & 66,573 &  7,397 &  49,271 &  5,474 & 10,000 &  4,195 \\
                          & 60/40 & 44,382 & 29,588 &  22,088 &  2,454 & 10,000 & 10,000 \\
\midrule
MS MARCO                  & 90/10 & 291,869 & 32,428 & 297,397 & 33,044 & 4,675 & 518 \\
\bottomrule
\end{tabular}
\label{tab:data_statistics}
\end{table*}

\subsection{Baseline details}\label{baselines}
\begin{enumerate}
\item \textbf{Hierarchical $k$-means Semantic IDs:} \textbf{DSI}~\citep{tay2022transformer} and \textbf{NCI}~\citep{wang2022neural}. Both assign identifiers by recursive $k$-means over document embeddings. The remaining difference between the two methods is architectural.

\item \textbf{Product Quantization Semantic IDs:} \textbf{ULTRON-PQ}~\citep{zhou2022ultron}, in which every identifier token is a quantized code of a document embedding.

\item \textbf{RQ and RQ-VAE Semantic IDs:} Residual quantization DocID is first used in \textbf{MEVI} \citep{zhang2023mevi}; RQ-VAE DocID is first used in generative recommendation by \textbf{TIGER}~\citep{rajput2023recommender}. We adopt the identifier under the DSI architecture so that it differs from the hierarchical $k$-means baselines only in how identifiers are constructed.
\item \textbf{Textual IDs:} In addition, we include two non-numeric baselines to contrast numeric identifiers with
text-derived ones: \textbf{ULTRON-URL}~\citep{zhou2022ultron}, which assigns each document a title- and URL-derived string, and \textbf{SEAL}~\citep{bevilacqua2022autoregressive}, which
assigns no identifier at all and instead generates document substrings resolved through an
FM-index.

\end{enumerate}
\subsection{Hyperparameters}\label{hyper_appen}
Table \ref{tab:hparams} shows the hyperparameters we used. We did not tune any training hyperparameters: a single configuration is used for every dataset and split, and only the checkpoint is selected by early stopping on  seen document queries from valdiation set.

Additionally, We train with additional pseudo-queries, following prior work. For NQ320K and TriviaQA, we use NCI's released pseudo -queries~\citep{wang2022neural}, generated by docTTTTTquery \citep{nogueira2019doc2query}, with 10 pseudo-queries per document in $D_0$.  The total training budget for our method is 96 A100 GPU-hours. Each reproduced baseline is trained with the same pseudo-queries and budget hours. For MS MARCO, we follow prior work~\citep{zhang2026model}, using up to 5 pseudo-queries per document in $D_0$.

\begin{table*}[t]
\centering
\small
\begin{tabular}{lllll}
\toprule
 & \multicolumn{2}{c}{BERT-base} & \multicolumn{2}{c}{T5-base} \\
\cmidrule(lr){2-3} \cmidrule(lr){4-5}
 & $M_1$ & $M_2$ & $M_1$ & $M_2$ \\
\midrule
Backbone            & BERT-base & from $M_1$ & T5-base encoder & from $M_1$ \\

Objective           & InfoNCE & InfoNCE over $D_0$ DocID & InfoNCE & InfoNCE over $D_0$ DocID \\
Temperature         & $\tau_1{=}0.07$ & $\tau_2{=}0.05$ ($\gamma{=}20$) & $\tau_1{=}0.07$ & $\tau_2{=}0.05$ ($\gamma{=}20$) \\
Optimizer           & AdamW & AdamW & Adafactor & Adafactor \\
Learning rate       & $2{\times}10^{-5}$ & $2{\times}10^{-5}$ (enc.) / $10^{-3}$ ($\mathbf{w}$) & $10^{-3}$ & $10^{-3}$ (enc.) / $10^{-3}$ ($\mathbf{w}$) \\
Effective batch     & $512{\times}2$ GPUs $=1{,}024$ & $256{\times}2$ accum.\ $=512$ & $512{\times}2$ GPUs $=1{,}024$ & $256{\times}2$ accum.\ $=512$ \\
Budget              & $\leq$200 epochs & $\leq$100k steps & $\leq$200 epochs & $\leq$100k steps \\
Early stopping      & patience 5 & patience 10 & patience 5 & patience 10 \\
\bottomrule
\end{tabular}
\caption{Training hyperparameters. PDIM: MRR of the gold document under exhaustive ranking by
bitwise agreement, without a prefix tree. $M_2$'s validation set is the official dev set for
NQ320K and TriviaQA and a held-out 10\% slice of the $D_0$ training queries for MS MARCO.}
\label{tab:hparams}
\end{table*}

\subsection{Retrieval result on $|D_0|{=}90\%$ seen / $|D_1|{=}10\%$ unseen split}\label{dyn10v2}
Table \ref{tab:dyn10v2} shows the result of $|D_0|{=}90\%$ seen / $|D_1|{=}10\%$ unseen split. The same findings from $|D_0|{=}60\%$ seen / $|D_1|{=}40\%$ can be seen.

\begin{table*}[t]
\centering
\small
\resizebox{0.9\textwidth}{!}{%
\begin{tabular}{llcccccccc}
\toprule
\multirow{2}{*}{Method} & \multirow{2}{*}{DocID type} & \multicolumn{4}{c}{NQ320K} & \multicolumn{4}{c}{TriviaQA} \\
\cmidrule(lr){3-6} \cmidrule(lr){7-10}
 & & Hit@1 & Hit@10 & Hit@100 & MRR@100 & Hit@1 & Hit@10 & Hit@100 & MRR@100 \\
\midrule
\rowcolor{gray!15}
\multicolumn{10}{l}{\textit{Queries with unseen target documents ($D_1$)}} \\
NCI & Hier. $k$-means & 0.09 & 0.49 & 2.03 & 0.24 & 0.00 & 0.12 & 0.62 & 0.03 \\
DSI & Hier. $k$-means & 0.09 & 0.64 & 3.95 & 0.30 & 0.01 & 0.23 & 1.57 & 0.08 \\
NCI (shared IDs) & Hier. $k$-means & 0.90 & 6.68 & 18.04 & 2.83 & 0.34 & 4.84 & 15.80 & 1.76 \\
DSI (shared IDs) & Hier. $k$-means & 0.67 & 5.34 & 14.11 & 2.22 & 0.38 & 3.97 & 13.23 & 1.59 \\
DSI & RQ-VAE  & 0.06 & 0.66 & 3.63 & 0.28 & 0.01 & 0.11 & 1.53 & 0.06 \\
ULTRON-PQ & Product quant. & 0.05 & 0.43 & 12.09 & 0.32 & 0.00 & 0.00 & 5.08 & 0.06 \\
ULTRON-URL & URL / title string & 14.80 & 31.25 & 46.53 & 20.60 & -- & -- & -- & -- \\
SEAL (BART) & Substrings & 15.05 & 58.88 & 83.13 & 29.47 & 27.60 & 58.62 & 78.71 & 38.11 \\
CoBIGR-Bert & Bit-factorized code & \textbf{45.02} & \textbf{79.10} & \textbf{85.83} & \textbf{57.73} & \textbf{45.70} & \textbf{73.73} & \textbf{82.29} & \textbf{55.95} \\
CoBIGR-T5 & Bit-factorized code & \underline{42.90} & \underline{77.48} & \underline{85.10} & \underline{55.60} & \underline{44.22} & \underline{70.35} & \underline{80.00} & \underline{53.60} \\
\midrule
\rowcolor{gray!15}
\multicolumn{10}{l}{\textit{Queries with seen target documents ($D_0$)}} \\
NCI & Hier. $k$-means & 66.15 & 85.85 & \textbf{92.71} & 73.46 & \underline{71.77} & \textbf{91.47} & \textbf{96.45} & \underline{79.31} \\
DSI & Hier. $k$-means & 61.44 & 77.49 & 85.07 & 67.60 & 63.17 & 83.10 & 89.99 & 70.62 \\
DSI & RQ-VAE  & 58.15 & 78.52 & 85.90 & 65.87 & 63.36 & 84.26 & 90.79 & 71.24 \\
ULTRON-PQ & Product quant. & 60.58 & 79.18 & 86.46 & 67.60 & 66.88 & 84.87 & 91.62 & 73.71 \\
ULTRON-URL & URL / title string & 63.71 & 79.36 & 87.04 & 69.78 & -- & -- & -- & -- \\
SEAL (BART) & Substrings & 62.99 & 83.44 & 92.61 & 70.53 & 71.45 & 85.92 & 92.74 & 76.86 \\
CoBIGR-Bert & Bit-factorized code & \textbf{69.37} & \textbf{87.01} & \underline{91.09} & \textbf{75.99} & \textbf{73.22} & \underline{89.75} & \underline{93.44} & \textbf{79.55} \\
CoBIGR-T5 & Bit-factorized code & \underline{68.33} & \underline{85.89} & 90.68 & \underline{74.92} & 70.79 & 88.45 & 92.44 & 77.41 \\

\bottomrule
\end{tabular}
}
\caption{Retrieval results on NQ320K and TriviaQA, $|D_0|{=}90\%$ seen / $|D_1|{=}10\%$ unseen split; all methods retrieve from the full index.}
\label{tab:dyn10v2}
\end{table*}

\subsection{Full corpus}\label{full_corpus}

\paragraph{Dataset} For the full-corpus evaluation, we follow the prior work: NQ320K contains 109,739 documents with 307,373 training and 7,830 validation queries (the test set was not released); TriviaQA contains 73,970 documents with 61,891 training, 7,993 validation, and 7,701 test queries.

\paragraph{Baslines}
On the full corpus, where no documents are added after indexing, besides the baselines used for the dynamic-corpus setting, we additionally compare against recent generative retrieval systems that report on this benchmark, including Learnable numeric DocID \textbf{GenRet}~\citep{sun2023learning}, and Text DocID \textbf{Novo}~\citep{wang2023novo}, we also report the none-generative search results as reference. 

\paragraph{Result}
Table~\ref{tab:full} shows the static setting ($|D_1| = 0$), where all documents are seen during training and generative retrieval baselines can make the most effective use of indexing training compared to the dynamic setting. Our method remains competitive by performing only query-to-DocID optimization: it achieves 69.11$\%$ Hit@1 on NQ320K and 75.51$\%$ Hit@1, and maintains comparable results on other metrics. Using more tokens improves performance, while latency is relatively insensitive to identifier length (23.7--27.7 ms for 7--21 tokens).

\begin{table*}[t]
\centering
\resizebox{0.92\textwidth}{!}{%
\begin{tabular}{lcccccccccc}
\toprule
\multirow{2}{*}{Method} & \multicolumn{5}{c}{NQ320K (Validation set)} & \multicolumn{5}{c}{TriviaQA} \\
\cmidrule(lr){2-6} \cmidrule(lr){7-11}
 & Hit@1 & Hit@10 & Hit@100 & MRR@100 & B/doc & Hit@1 & Hit@10 & Hit@100 & MRR@100 & B/doc \\
\midrule
DSI$^\dagger$ & 60.87 & 76.90 & 84.37 & 66.94 & 3.0 & 63.38 & 84.09 & 91.16 & 70.99 & 2.5 \\

SEAL$^\dagger$ & 61.38 & 82.96 & 93.05 & 69.39 & 3{,}134 & 70.71 & 86.85 & 93.30 & 76.76 & 1{,}901 \\
NCI$^\dagger$ & 65.90 & 85.06 & 92.80 & 73.04 & 3.0 & 72.78 & \textbf{92.61} & 97.29 & 80.35 & 2.5 \\
ULTRON-PQ$^\dagger$ & 58.90 & 78.44 & 86.19 & 66.34 & 24 & 66.63 & 86.11 & 92.70 & 74.08 & 24 \\
ULTRON-URL$^\dagger$ & 62.05 & 79.31 & 86.85 & 68.74 & 27 & -- & -- & -- & -- & -- \\
GenRet \citep{sun2023learning} & 68.1 & 88.8 & 95.2 & 75.9 & -- & -- & -- & -- & -- & -- \\
MEVI (full ensemble)$^\dagger$ & 63.04 & 84.55 & 95.27 & 71.55 & 3{,}074.5 & 70.94 & 90.65 & \textbf{97.35} & 78.60 & 3{,}074.5 \\
NOVO \citep{wang2023novo} & \textbf{69.3} & \textbf{89.7} & \textbf{95.9} & \textbf{76.7} & -- & -- & -- & -- & -- & -- \\
\midrule
CoBIGR-T5 & 67.71 & 86.00 & 90.55 & 74.70 & 15.0 & 71.24 & 89.72 & 93.64 & 78.32 & 15.0 \\
CoBIGR-Bert & 68.46 & 87.04 & 91.57 & 75.50 & 15.0 & \textbf{74.08} & 91.04 & 94.34 & \textbf{80.63} & 15.0 \\
\midrule
\rowcolor{gray!15}
\multicolumn{11}{l}{\textit{full-corpus search}} \\
BERT-FT & 65.25 & 88.45 & 95.12 & 73.91 & 3{,}072 & 69.86 & 91.44 & 96.84 & 78.12 & 3{,}072 \\
GTR-T5-base & 54.94 & 83.70 & 93.60 & 65.30 & 3{,}072 & 60.16 & 83.76 & 93.46 & 68.70 & 3{,}072 \\
BM25 & 36.93 & 70.28 & 88.31 & 48.41 & 1{,}568 & 51.94 & 81.48 & 92.87 & 62.70 & 1{,}037 \\
\bottomrule
\end{tabular}
}
\caption{Retrieval results on the official splits. All documents are seen documents ($|D_1| = 0$). ``--'' means the metric is not reported in the original paper.
$^\dagger$Our reproduction. BERT-FT refers to fine-tuned on the NQ and TriviaQA datasets.
\textbf{B/doc}: The bytes the index must store per document.
}
\label{tab:full}
\end{table*}

\subsection{Efficiency analysis}\label{pareto_hit1}
\begin{figure*}[t]
        \centering
        \includegraphics[width=0.75\textwidth]{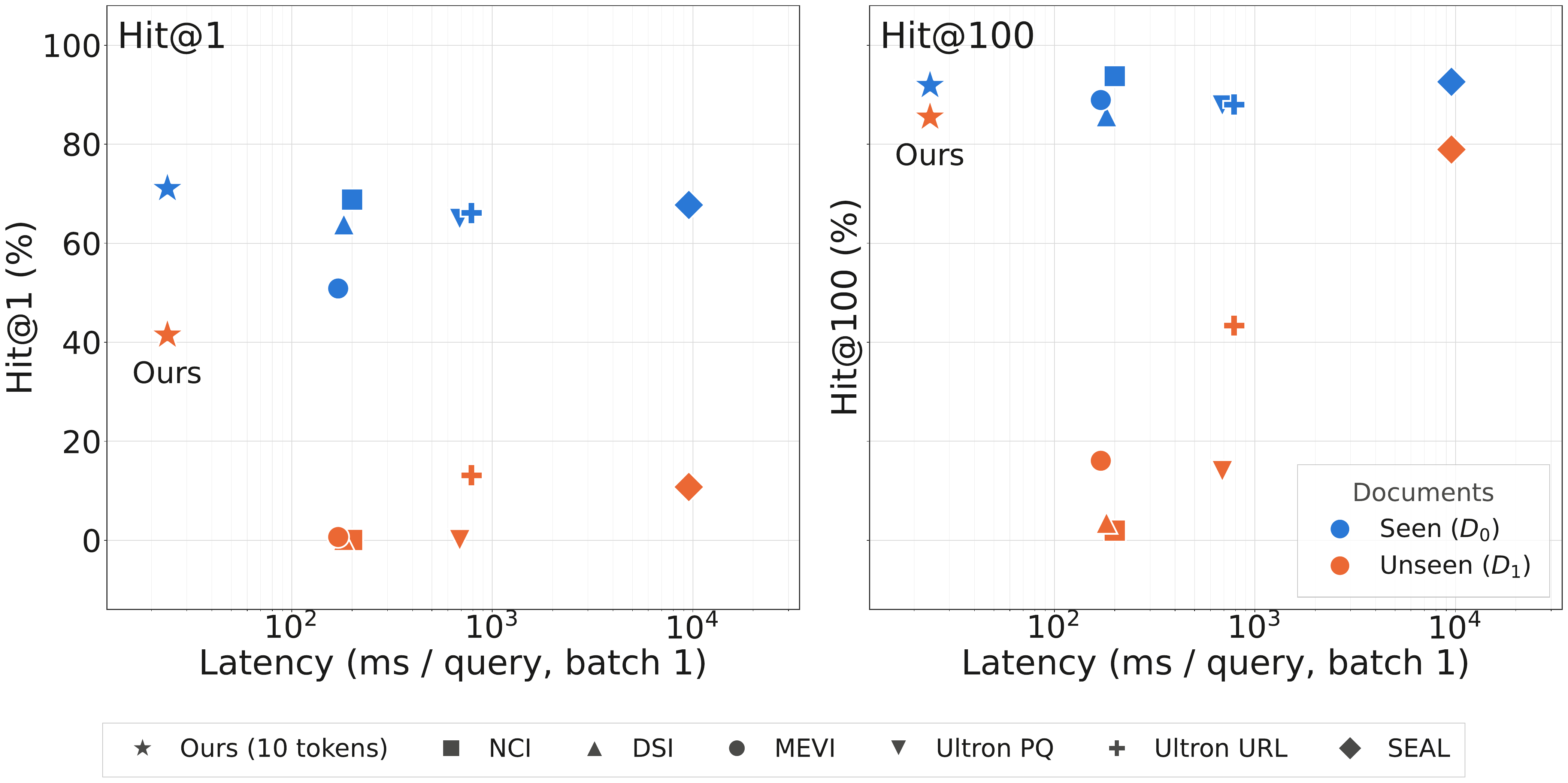}
        \caption{Latency against Hit on the NQ dataset, batch size of 1 and 1 GPU. ($|D_0|{=}60\%$, $|D_1|{=}40\%$)}
        \label{fig:pareto_hit1}
\end{figure*}
Figure~\ref{fig:pareto_hit1} shows the latency against hit rate. On unseen documents, our method achieves the highest Hit@1 and Hit@100 while also being the most efficient. SEAL generalizes relatively well to unseen documents, but retrieving the documents matching its candidate substrings through the FM-index introduces substantial query-time overhead. On seen documents, our method achieves the highest Hit@1, while NCI achieves the highest Hit@100; even so, our method remains substantially faster than all baselines because the score of our method decomposes over bits and does not condition on previously generated tokens, the constrained search that follows requires no more forward pass.

\subsection{Effect of multiple trees- hit@1 and hit@100} \label{m_tree}
\begin{figure*}[t]
        \centering
        \includegraphics[width=0.75\textwidth]{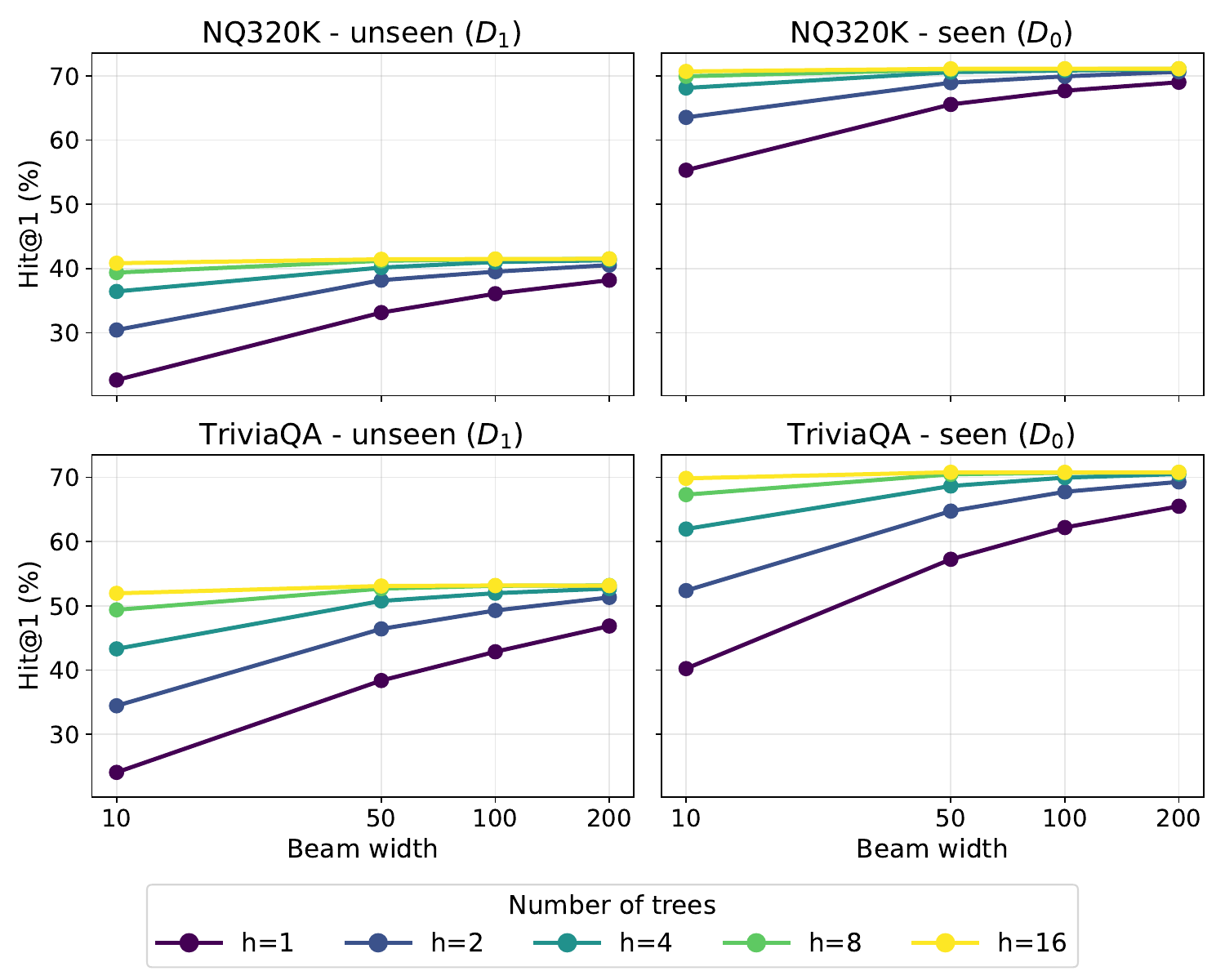}
        \caption{Hit@1 by different size of trees and beams ($|D_0|{=}60\%$, $|D_1|{=}40\%$) }
        \label{fig:R1_beam_tree}
\end{figure*}

\begin{figure*}[t]
        \centering
        \includegraphics[width=0.75\textwidth]{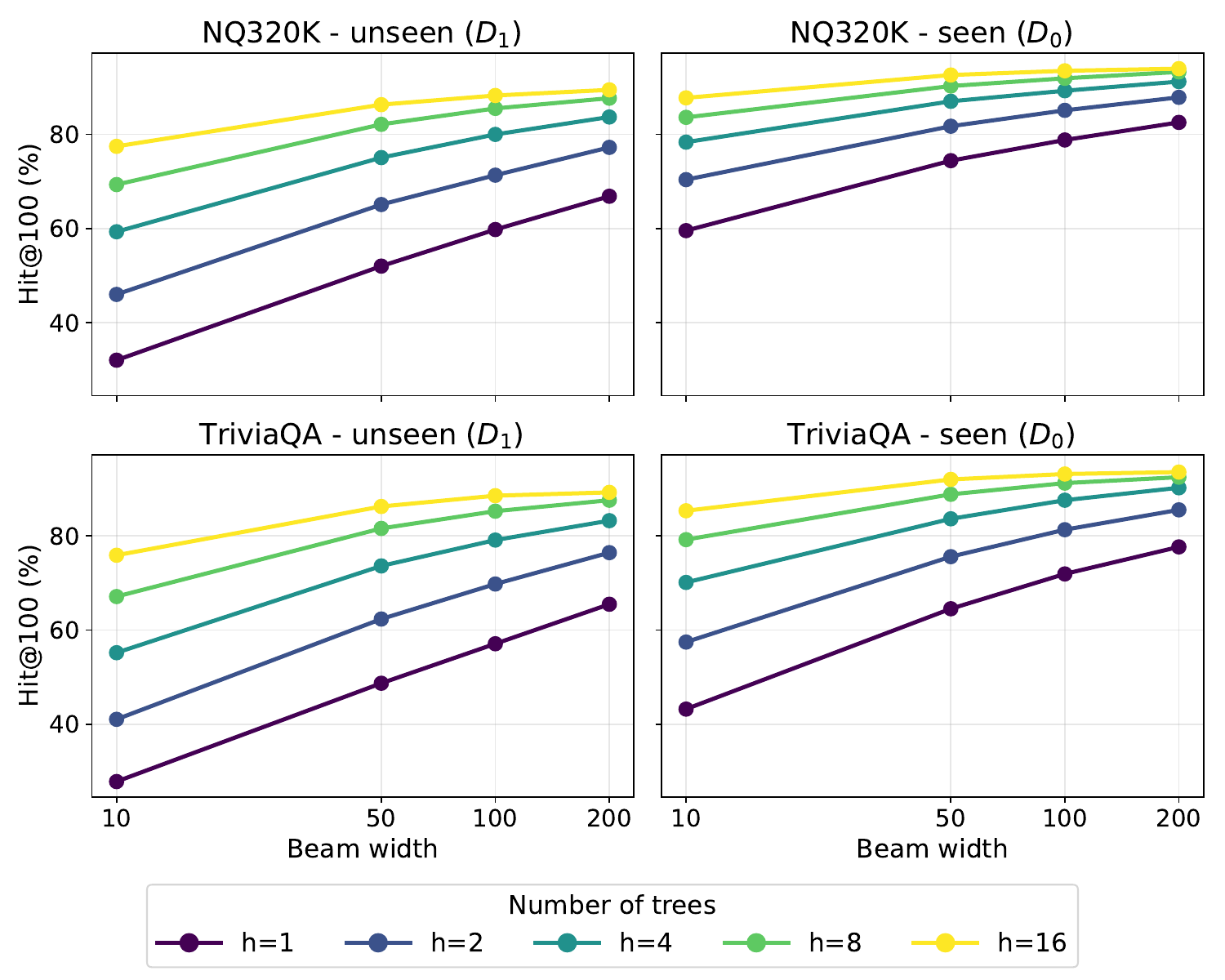}
        \caption{Hit@100 by different size of trees and beams ($|D_0|{=}60\%$, $|D_1|{=}40\%$)}
        \label{fig:R100_beam_tree}
\end{figure*}

Figures \ref{fig:R1_beam_tree} and \ref{fig:R100_beam_tree} show the simialr result as hit@10 in Figure \ref{fig:R10_beam_tree}.

\end{document}